\documentclass[fleqn,usenatbib]{mnras}

\usepackage{newtxtext,newtxmath}

\usepackage[T1]{fontenc}

\DeclareRobustCommand{\VAN}[3]{#2}
\let\VANthebibliography\thebibliography
\def\thebibliography{\DeclareRobustCommand{\VAN}[3]{##3}\VANthebibliography}

\usepackage{graphicx}	
\usepackage{amsmath}	
\usepackage{xcolor, color}

\newcommand{\msun}{\,${\rm M_\odot}$}

\definecolor{deeporange}{rgb}{0.91, 0.41, 0.17}

\title[AGN feedback in SMUGGLE]{AGN feedback in isolated galaxies with a
SMUGGLE multiphase ISM}

\author[Sivasankaran et al.]{
Aneesh Sivasankaran$^{1}$\thanks{E-mail: aneeshs@ufl.edu},
Laura Blecha$^{1}$,
Paul Torrey$^{2}$,
Luke Zoltan Kelley$^{3}$,
Aklant Bhowmick$^{1}$,\newauthor
Mark Vogelsberger$^{4}$,
Lars Hernquist$^{5}$,
Federico Marinacci$^{6,7}$ and
Laura V. Sales$^{8}$
\\
$^{1}$Department of Physics, University of Florida, Gainesville, Florida 32601, USA\\
$^{2}$Department of Astronomy, University of Virginia, 530 McCormick Road, Charlottesville, VA 22903, USA\\
$^{3}$Department of Astronomy, University of California at Berkeley, 501 Campbell Hall, Berkeley, CA 94720, USA\\
$^{4}$Department of Physics, Kavli Institute for Astrophysics and Space Research, Massachusetts Institute of Technology, Cambridge, MA 02139, USA\\
$^{5}$Harvard-Smithsonian Center for Astrophysics, 60 Garden Street, Cambridge, MA 02138, USA\\
$^{6}$Department of Physics \& Astronomy ``Augusto Righi'', University of Bologna, via Gobetti 93/2, 40129 Bologna, Italy\\
$^{7}$INAF, Astrophysics and Space Science Observatory Bologna, Via P. Gobetti 93/3, I-40129 Bologna, Italy\\
$^{8}$University of California, Riverside, 900 University Ave., Riverside, CA 92521, USA\\
}

\date{Accepted XXX. Received YYY; in original form ZZZ}

\pubyear{2024}

\begin{document}
\label{firstpage}
\pagerange{\pageref{firstpage}--\pageref{lastpage}}
\maketitle

\begin{abstract}
Feedback from active galactic nuclei (AGN) can strongly impact the host galaxies by driving high-velocity winds that impart substantial energy and momentum to the interstellar medium (ISM). In this work, we study the impact of these winds in isolated galaxies using high-resolution hydrodynamics simulations. Our simulations use the explicit ISM and stellar evolution model called Stars and MUltiphase Gas in GaLaxiEs (SMUGGLE). Additionally, using a super-Lagrangian refinement scheme, we resolve AGN feedback coupling to the ISM at $\sim$10-100 pc scales. We find that AGN feedback efficiently regulates the growth of SMBHs. However, its effect on star formation and outflows depends strongly on the relative strengths of AGN vs local stellar feedback and the geometrical structure of the gas disk. When the energy injected by AGN is subdominant to that of stellar feedback, there are no significant changes in the star formation rates or mass outflow rates of the host galaxy. Conversely, when the energy budget is dominated by the AGN, we see a significant decline in the star formation rates accompanied by an increase in outflows. Galaxies with thin gas disks like the Milky Way allow feedback to escape easily into the polar directions without doing much work on the ISM. In contrast, galaxies with thick and diffuse gas disks confine the initial expansion of the feedback bubble within the disk, resulting in more work done on the ISM. Phase space analysis indicates that outflows primarily comprise hot and diffuse gas, with a lack of cold and dense gas.
\end{abstract}





\section{Introduction}

It is widely accepted that supermassive black holes (SMBHs) exist at the centers of the majority of massive galaxies. These BHs are thought to start as $\sim 10^2-10^5$\msun\  seeds and grow in size by many orders of magnitude primarily through gas accretion \citep{Soltan1982}. During accretion, a fraction of the accreted energy is converted into radiation resulting in luminosities as high as $\sim 10^{48}$ergs s$^{-1}$ emanating from milliparsec size regions. These objects are called Active Galactic Nuclei (AGNs)  and the coupling of this energy to the surrounding interstellar medium (ISM) of the host galaxy is called AGN feedback. AGN feedback is postulated to play a major role in the evolution of the host galaxy by regulating star formation by expelling cold gas or heating up gas. AGN feedback is also thought to self-regulate the growth of SMBHs by suppressing gas inflows. These postulates are motivated by observational data which show strong correlations between the mass of the SMBHs and the properties of their host galaxies such as bulge mass, stellar velocity dispersion and luminosity \citep{Kormendy1995,Magorrian1998,Ferrarese2000,Gultekin12009,mcconnell2013revisiting,kormendy2013coevolution,ReinesVolonteri2015scalingrelations,bennert2015scaling,savorgnan2016scaling}.
Apart from these correlations, there are also observations of strong outflows (with outflow rates $\sim1000\,$\msun yr$^{-1}$ and velocities $>1000\,$km s$^{-1}$) emanating from galactic nuclei \citep{Rupke2011,Sturm2011,cicone2014massive,fiore2017agn,fluetsch2019cold}, 
supporting the existence of AGN feedback and its importance in the evolution of the host galaxy. Nevertheless, the role of AGN feedback in galaxy evolution is still a debated topic due to the uncertainties in our understanding of how the feedback actually couples to the interstellar medium (ISM). This coupling can happen through various mechanisms such as jets, winds and radiation pressure. Presently, the contributions of these distinct modes and their corresponding coupling efficiencies remain poorly constrained.

Theoretical studies of galaxy formation and evolution using numerical simulations have been very successful at reproducing many observational data \citep[e.g.,][]{vogelsberger2014a,vogelsberger2014b,genel2014,schaye2015eagle,mcalpine2017eagle,tng-results1,tng-results2,dave2019simba}. AGN feedback is an essential component in these simulations, and it has been found to be necessary for regulating star formation in massive galaxies, particularly at low redshifts \citep{teyssier2011mass,dubois2013agn,sijacki2015,TngSmbhFeedback2018}. However, some studies suggest that the observed correlations between BH and galaxy properties can be explained by a common gas supply fueling the BHs and star formation \citep{angles2013black,angles2015torque,angles2017torquemodel}. Hence, more work is needed in this area to fully understand the role of AGNs in galaxy evolution.

The biggest challenge in modelling AGN feedback in these simulations is the large range of scales involved: from the accretion disks around the SMBHs at sub-parsec scales to the galactic outflows which happen at kpc to Mpc scales. 
In cosmological simulations, because of resolution limitations, AGN feedback is typically injected as thermal or kinetic 
energy directly into the ISM at $\sim$kpc scales without resolving the actual mechanism that heats up the gas. These crude modeling of such an important important physical mechanisms generates uncertainties that limit the predictive power of the simulations.

In this work, we use high-resolution idealized simulations of AGN feedback in a range of galactic environments, with the aim of resolving the scale at which the feedback first couples to the surrounding ISM. The AGN feedback mode we consider in this work is fast nuclear winds with velocities of the order of $\sim0.1c$. This type of winds can be driven by radiation from the accretion disks around SMBHs and can result in broad absorption lines in the UV or X-ray \citep{weymann1981absorption,turnshek1988bal,chartas2003xmm}. Such high energy winds can have a significant impact on the galaxy; for example by shock-heating the surrounding ISM and sweeping up matter into large-scale outflows. The amount of work done by the wind on the ISM depend on several factors, such as the geometry and porosity of the ISM \citep{wagner2012agncoupling,wagner2013ultrafast,bieri2017outflows,Torrey2020}. Feedback is expected to do more work on a dense and uniform ISM compared to a porous one, which can allow the hot gas to escape through low density channels \citep{faucherQuataert2012physics,Torrey2020}.

Therefore, an explicit ISM model that can accurately resolve the multiphase structure, geometry and density variations of the ISM is crucial for accurate studies of AGN feedback. In this work we employ the recently developed Stars and MUltiphase Gas in GaLaxiEs (SMUGGLE, \citealt{smuggle-paper}) model for the moving mesh hydrodynamics code AREPO \citep{Springel2010}. SMUGGLE is an explicit and comprehensive stellar feedback model which has been shown to produce an ISM with well resolved multiphase structure,  observationally consistent star formation rates, galactic outflows \citep{smuggle-paper}, H$\alpha$ emission line profiles \citep{tacchella2022h,smith2022physics} and star cluster properties \citep{HuiLi2020effects,HuiLi2021formation}, and constant density cores in dwarf galaxies \citep{jahn2021real}. In \citet{sivasankaran2022simulations} (Paper I hereafter) we analyzed BH accretion (without including AGN feedback) in isolated and merging galaxies with the SMUGGLE ISM. We showed that the gas turbulence in the ISM leads to highly variable accretion rates, which is qualitatively different from the way BHs accrete in simulations with widely used effective equation-of-state models of the ISM.

There are a few other explicit ISM models in literature such as the Feedback In Realistic Environments (FIRE; \citealt{FIRE-2}) framework in the meshless magnetohydrodynamics code GIZMO \citep{hopkins2015gizmo} and the \cite{Agertz2013} ISM in the adaptive mesh refinement code RAMSES \citep{teyssier2002ramses}. Recently, the FIRE model has been used to perform ultra high resolution ($\lesssim10$pc) but short timescale (10-50 Myr) studies of SMBH fueling \citep{anglesalcazar2021hyperLagrangian} and AGN feedback \citep{cochrane2023impact,mercedes2023local} in cosmological zoom-in simulations.

In this paper we implement a novel prescription for AGN feedback in the SMUGGLE framework in the simulations presented in Paper I. Our study aims to investigate the impact of feedback on the growth of SMBHs, generation of gas outflows, and star formation in isolated galaxies over $\sim$Gyr timescales . Additionally, we aim to explore how the properties of the host galaxy, such as the geometry of the disc and its gas fraction, affect the extent of feedback's impact on the galaxy.

In addition to having an explicitly modelled ISM, very high resolution is also needed to accurately resolve the coupling of AGN feedback to the ISM, which is expected to happen at sub-parsec to tens of parsec scales. However, running galactic or cosmological scale simulations at such high resolutions is generally not practical. To address this issue, we implemented a super-Lagrangian refinement scheme in Paper I, which increases the gas mass resolution in a small region around the BHs. By utilizing this method in lower-resolution simulations, we were able to reproduce the central gas dynamics and accretion rates of very high, uniform resolution simulations. This refinement scheme not only improves the estimation of accretion rates, but also enables us to resolve the initial coupling of AGN feedback to the ISM at scales as close to the radius of influence of the BH as possible by capturing finer gas density variations around the BHs. Similar refinement techniques have been applied in other studies to investigate BH accretion and feedback \citep{Curtis2015,anglesalcazar2021hyperLagrangian}.

This paper is structured as follows: In Section \ref{section:Methods}, we provide a brief overview of the simulation code AREPO, the SMUGGLE model, the BH accretion and feedback subgrid models, and the super-Lagrangian refinement scheme. Section \ref{section:results} presents the findings from our isolated galaxy simulations. Finally, in Section \ref{section:discussions&conclusions}, we discuss our results and draw our conclusions.

\section{Methods}\label{section:Methods}
\subsection{AREPO and SMUGGLE overview}
The simulations in the paper are carried out using the moving-mesh magnetohydrodynamics code AREPO \citep{Springel2010,pakmor2016improving}. AREPO solves magnetohydrodynamics equations on an unstructured Voronoi mesh using a finite volume method. The mesh points are allowed to move along with the fluid flow allowing an automatic adjustment of spatial resolution. 
AREPO models dark matter and stars as collisionless particles obeying the Vlasov equation coupled to Poisson's equations (gravity). The ISM is treated as an ideal gas following 
Euler's equations. The ISM physics, such as star formation, stellar feedback, gas heating and cooling, are modelled explicitly in our simulations using the SMUGGLE model \citep{smuggle-paper}. 

In SMUGGLE, gas heating and cooling includes mechanisms such as cosmic rays, photoelectric heating, and low temperature atomic and molecular cooling, which can create a multiphase ISM with temperatures from $\sim10$K to $\sim10^8$K. Star formation and stellar feedback are implemented locally in SMUGGLE. Star particles are allowed to form only from cold, dense and gravitationally bound gas cells. A density threshold of 100 cm$^{-3}$ is adopted for star formation in our simulations. Stellar feedback (including supernova energy and momentum, radiative feedback from stars, and AGN and OB wind momentum) is injected into the nearest gas cells of the star particles in a kernel-weighted fashion. All stellar feedback inputs are calculated by sampling individual stars from the star particles using the \citet{ChabrierIMF} IMF. We refer the reader to \citet{smuggle-paper} for more details of the included physical processes and their implementation.

\begin{figure*}
    \centering
    \includegraphics[width=\textwidth]{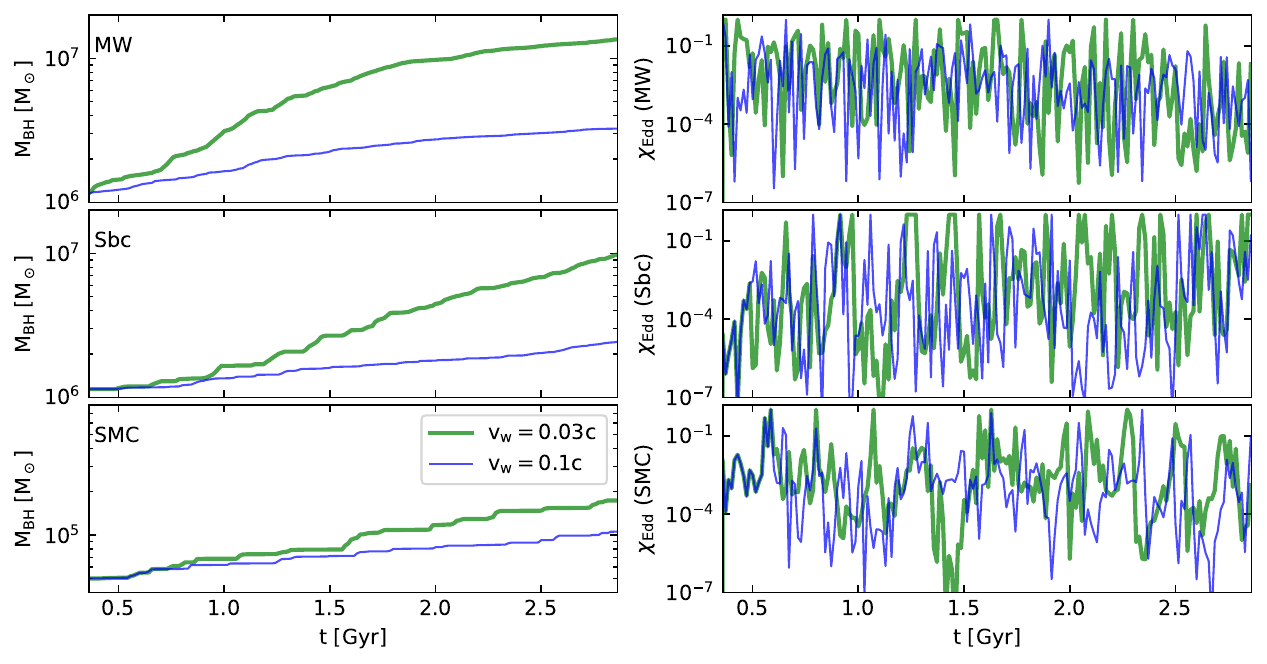}
    \caption{BH mass (left) and Eddington ratio (right) as functions of time for the MW, Sbc and SMC isolated galaxy runs with wind velocities of $0.1c$ and $0.03c$. BH mass growth is effectively regulated by the fast AGN winds in all galaxies. $\rm{v_w=0.1c}$ results in a factor of $2-3$ BH mass growth over 2.5 Gyr in all galaxies. Lowering the wind velocity leads to higher accretion rates and higher BH masses. Eddington ratios have large fluctuations due to transient cavities created primarily by bursty stellar feedback. Most of the BH growth happens during the peak Eddington ratios. When the wind velocity is lowered, the BHs spend more time in the high accretion state.}
    \label{fig:bhmass_eddratio_vs_vel}
\end{figure*}

\subsection{Black Hole Subgrid Models}
\subsubsection{BH Accretion}
In our simulations that include AGN feedback, BH accretion rates are calculated based on the Eddington-limited Bondi-Hoyle prescription. In this approach the gas around the BH is assumed to be spherically symmetric and the accretion rate is estimated from the density and temperature of the gas as given below:
\begin{equation}\label{eqn:mdot}
    \dot{M}_{\rm BH}=\mathrm{min}(\dot{M}_{\mathrm{Bondi}},\dot{M}_{\mathrm{Edd}}),
\end{equation}
where $\dot{M}_{\mathrm{Bondi}}$ is the Bondi-Hoyle accretion rate,
\begin{equation}\label{eqn:bondi-rate}
    \dot{M}_{\mathrm{Bondi}}=\frac{4\pi G^2 M^2_{\rm BH}\rho}{c^3_s},
\end{equation}
and $\dot{M}_{\mathrm{Edd}}$ is the Eddington limit,
\begin{equation}\label{eqn:eddington limit}
    \dot{M}_{\mathrm{Edd}}=\frac{4\pi G M_{\rm BH} m_p}{\epsilon_r \sigma_T c},
\end{equation}
which represents the accretion rate at which the radiation pressure becomes equal to the gravitational pull on the gas. In Equations \ref{eqn:bondi-rate}-\ref{eqn:eddington limit}, $\rho$ and $c_s$ are the kernel-weighted (see Paper I for details) gas density and sound speed of the gas near the BH, $G$ is Newton's constant, $c$ is the speed of light, $\epsilon_r$ is the radiative efficiency, which is set to 0.1, $M_{\rm BH}$ is the mass of the BH, $m_p$ is the mass of the proton and $\sigma_T$ is the Thompson cross section. We will often use the Eddington ratio, defined as $\chi_{\rm Edd} \equiv \dot M_{\rm BH}/\dot M_{\rm Edd}$, 
to parameterize the accretion rates. 

In addition to our fiducial simulations in which $\dot M_{\rm BH}$ is calculated from the BH and gas properties according to Equations \ref{eqn:mdot}-\ref{eqn:eddington limit}, we also run a set of simulations with the accretion rate fixed at all times to 20\% of the initial Eddington accretion rate. While artificial, this subset of simulations helps  to disentangle the impacts of AGN feedback on the galaxy from the impacts of AGN feedback on the BH fueling rate. These results are described in \S\ \ref{ssec:fixed_mdot}.

\subsubsection{AGN feedback}
In this work we model AGN feedback via fast nuclear winds, which are assumed to be driven by physics happening on scales unresolved in our simulations such as the accretion disc and broad line region. In our model we assume that a fraction of the AGN luminosity drives fast kinetic winds with velocities of the order of $\sim 0.1c$ radially away from the BH. Our model involves two free parameters which prescribe the strength of the wind: (i) the wind velocity ($v_{\rm w}$) and (ii) the wind mass loading factor $(\eta_{\rm w})$, which is the ratio of the wind mass outflow rate $(\Dot{M}_{\rm w})$ and the BH growth rate $(\Dot{M}_{\rm BH})$. At each time-step the wind mass is calculated as $M_w = \eta_w \Dot{M}_{\rm BH} \Delta t$, and the gas cells around the BHs are kicked stochastically. Specifically each gas cell in the BH's kernel is kicked with a velocity of $v_w$ in a random direction with a probability $p=M_w/M_{\rm kernel}$ where $M_{\rm kernel}$ is the total mass of all gas cells in the kernel. This ensures energy and momentum conservation in a time-averaged sense. We inject feedback only when the Eddington ratio is above 0.05. 

Note that it is possible to put a mass and(or) distance dependent weight in the kick probability, but we find that this can lead to a slight (factor of few) violation of energy conservation. Thus, we keep the probability uniform inside the kernel. It is also possible to inject the feedback energy in a continuous manner at every time-step, but this method might not be able to capture the effects of the shocks created by the time-varying, high-velocity wind with the surrounding gas. Note than $v_w\sim 0.1c$ is roughly consistent with the quasar-mode feedback coupling efficiency assumed in literature. A coupling efficiency of $5\%$ results in a velocity of $0.15c$.

We also simulate each of the three isolated galaxies with no AGN feedback, to better quantify the impact of AGN feedback on the galactic environment. In these simulations, the BHs are not allowed to accrete at all and act merely as passive particles. This avoids unphysically high accretion rates in the absence of AGN feedback, an issue described in Paper I and in Section~\ref{section:results} below.

\subsubsection{Super-Lagrangian refinement}
We recently implemented a super-Lagrangian refinement scheme in AREPO to accurately resolve accretion rates and gas dynamics near the BH in low resolution simulations \citep{sivasankaran2022simulations}. We use the same refinement scheme in all the simulations in this work. In this method, the target gas mass in the simulation is varied as a function of distance from the BH as follows:
\begin{align}
    m(r)= \begin{cases} 
      \displaystyle\frac{m_0}{F} & r\leq r_{\rm min} \\
      \displaystyle\frac{m_0}{F}\left(1+(F-1)\frac{r-r_{\rm min}}{r_{\rm max}-r_{\rm min}}\right) & r_{\rm min}< r\leq r_{\rm max} \\
      \displaystyle m_0 & r_{\rm max}< r \label{refinement equation}
        \end{cases}
\end{align}
where r is the distance of the gas cells from the nearest BH, $m_0$ is the uniform target mass, and $F$ is the refinement factor. The two radii $r_{\rm min}$ and $r_{\rm max}$ are set to 2.86 kpc (= 2 kpc/h) and 14.3 kpc (=10 kpc/h), and the refinement factor $F$ is set to 10 in all our simulations. For more details the reader is referred to \citet{sivasankaran2022simulations}. Gas cells in the refinement region of our fiducial simulations have a median size of $\sim30\,$pc and a minimum softening length of 45 pc. The kernel radius of the BHs with the weighted number of neighbors set to 64 is $\sim100\,$pc.

\begingroup
\setlength{\tabcolsep}{4pt}
\begin{table}
    \centering
    \begin{tabular}{|l|r|r|r|r|r|}\hline \hline
        Name &${\rm M_{200}}$ & ${\rm M_{BH}}$ & ${\rm M_{bulge}}$ & ${\rm M_{disk}}$ & Disk gas \\
         & $({\rm M_\odot})$ & $({\rm M_\odot})$ & $({\rm M_\odot})$ & $({\rm M_\odot})$ & fraction\\ \hline
        MW & $1.43\times10^{12}$ & $1.14\times10^6$ &  $1.50\times10^{10}$ & $4.73\times10^{10}$ & 0.16\\
        Sbc & $2.14\times10^{11}$ & $1.14\times10^6$ & $1.43\times10^{9}$ & $5.71\times10^{9}$ & 0.59\\
        SMC & $2.86\times10^{10}$ & $5.00\times10^4$ & $1.43\times10^{7}$ & $1.86\times10^{8}$ & 0.86\\ \hline\hline
    \end{tabular}
    \caption{Initial parameters for the isolated galaxy simulations. Columns 2 - 6 show the total mass of the galaxy, initial BH mass, bulge mass, disk mass and disk gas fraction of the three isolated galaxy initial conditions. These initial conditions are the same as the ones used in \citet{sivasankaran2022simulations}.}
    \label{tab:IC_params}
\end{table}
\endgroup

\begin{figure*}
    \centering
    \includegraphics[width=\textwidth]{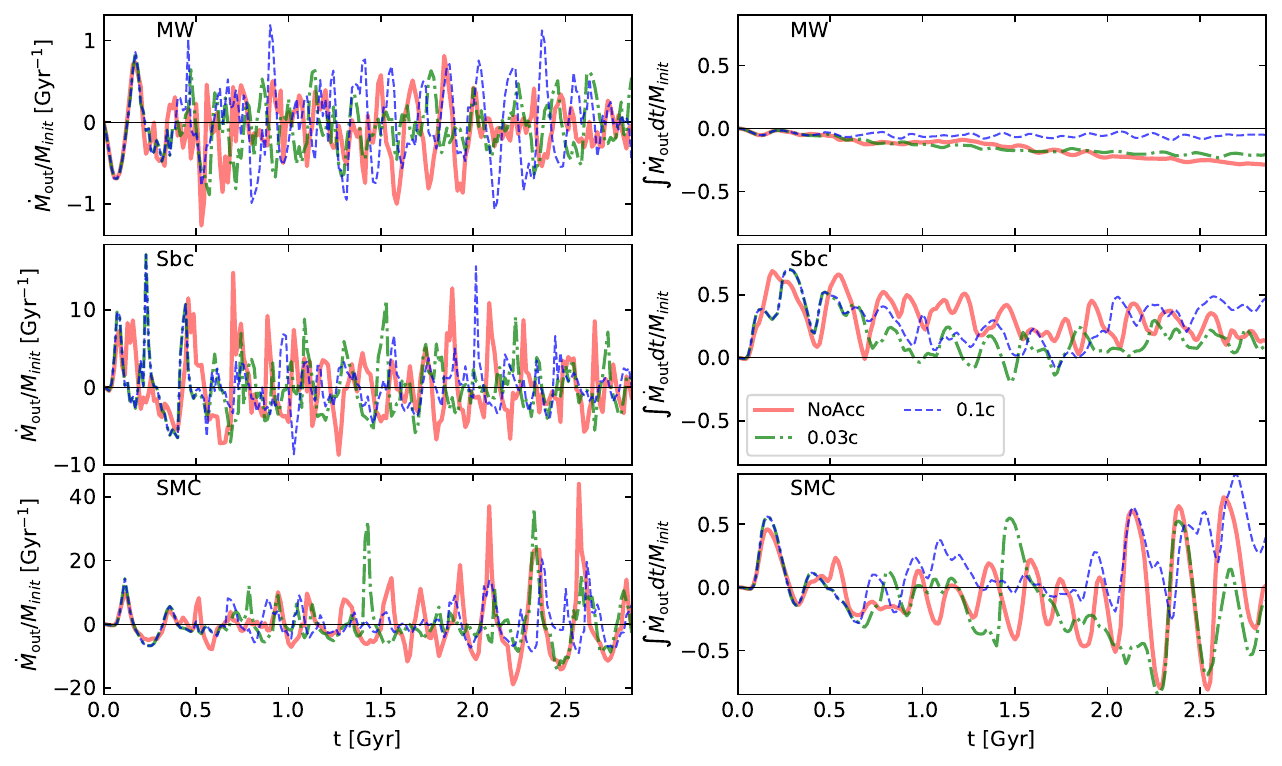}
    \caption{Instantaneous mass outflow rates (left) and cumulative mass outflow rates (right) normalized by the initial gas mass within the stellar half mass radius for MW (top), Sbc (middle) and SMC (bottom) galaxies. Each panel shows the time evolution for runs with wind velocities set to 0.1$c$, $0.03c$ and without BH feedback. Note that y-axis scales are different on the three panels on the left. With $\mathrm{v_w=0.1c}$, outflows are strongest in SMC. MW has slightly weaker outflows than SMC whereas Sbc shows no difference with feedback. None of the galaxies show significant impact in the case of $\mathrm{v_w=0.03c}$ 
    }
    \label{fig:outflowrates}
\end{figure*}

\subsection{Initial conditions}\label{section:IC}
We employ the same three different initial conditions (ICs) used in \citet{sivasankaran2022simulations} in our isolated galaxy simulations: A Milky-Way like galaxy (MW), a Small Magellanic Cloud like dwarf galaxy (SMC) and a luminous infrared type galaxy (Sbc). All three initial conditions contain a dark matter halo, stellar and gaseous disks, a stellar bulge  and a central BH. The parameters of the ICs are given in Table \ref{tab:IC_params}. The stellar bulge and dark matter halo follow the \citet{hernquist1990analytical} profile and the gaseous and stellar disks follow an exponential radial profile. The vertical structure of the gaseous disk is set by hydrostatic equilibrium and the stellar disk has a ${\rm sech}^2(z)$ profile. In all our simulations we set the (unrefined) target gas mass to $2.5\times10^4$\msun, dark matter particle mass to $2.64\times10^5$\msun, stellar bulge particle mass to $2.59\times10^4$\msun, and stellar disk particle mass to $2.78\times10^4$\msun. We also use the super-Lagrangian refinement with a refinement factor of 10 in all the runs at all times. This resolution is the same as the resx0.1F10 resolution level in \citet{sivasankaran2022simulations}. We also allow an initial relaxation period of 0.36 Gyr for the star formation and BH accretion to reach an approximate steady state. BH accretion and feedback are turned on only after this period. The gas disks have scale heights of 0.31, 1.34 and 1.07 kpc and scale radii of 21.0, 8.6 and 6.9 kpc for MW, Sbc and SMC respectively. We define these scale lengths as the distances at which the gas density decreases by a factor of 10 after the disks have reached steady state. Note that although Sbc and SMC are much smaller compared to the overall size of MW, their gas disks are much thicker than that of the MW. We refer to this set of three simulations as our fiducial simulation suite, the results of which are discussed below in \S~\ref{ssec:fiducial}. We separately simulate the same three galaxies with a fixed accretion rate and present these results in \S~\ref{ssec:fixed_mdot}. We also run the fiducial simulations with a refinement factor of 30 and also with a 10 times higher uniform resolution to test resolution convergence.

\section{Results}\label{section:results}
Our study involves the simulations of three galaxies with and without AGN feedback. In the runs without feedback, BHs are not allowed to accrete gas from the ISM. The feedback runs consist of two sets of simulations, each with different wind velocities ($v_w=0.1c$ and $0.03c$). Additionally, we also perform the $v_w=0.1c$ feedback runs with fixed accretion rates (20\% of initial Eddington rate). In subsequent sections, we examine the effects of the wind feedback on each of the three galaxies.

\subsection{Fiducial simulation suite}
\label{ssec:fiducial}
\subsubsection{Black hole mass growth and Eddington ratios}
In the left panels of Figure \ref{fig:bhmass_eddratio_vs_vel}, we compare the BH mass growth in the MW, Sbc, and SMC isolated galaxy runs with wind velocities of $0.1c$ and $0.03c$. We can see that feedback with a wind velocity of $v_w=0.1c$ effectively regulates BH growth, resulting in  moderate BH mass growth in all three galaxies over a period of $2.5\,$Gyr. The BHs in the MW, Sbc, and SMC grow by factors of 2.85, 2.13, and 2.12, respectively. With a lower AGN wind velocity of 0.03c, BH growth is more efficient, especially in MW and Sbc. The BHs grow by factors of 11.9 and 8.57 in these galaxies, respectively, over the same period, while the BH mass in SMC grows by a factor of 3.48. While this may provide a hint that BH growth in SMC is more efficiently regulated by feedback than in the other two galaxies, it also reflects the $M_{\rm BH}^2$ scaling of the Bondi accretion rate, given the lower initial BH mass in the SMC galaxy. In the following sections we explore the balance between accretion, feedback, and galactic environment in more detail.

In the right panels of Figure \ref{fig:bhmass_eddratio_vs_vel}, we plot the Eddington ratio of the BHs as a function of time. In all three galaxies, the Eddington ratio fluctuates by many orders of magnitude over few-Myr timescales. In Paper I we showed that bursty stellar feedback can produce transient cavities in the ISM that lead to these large fluctuations in background gas density and corresponding BH fueling rates. Figure \ref{fig:bhmass_eddratio_vs_vel} shows that this bursty stellar feedback strongly influences BH fueling rates even in the presence of AGN feedback. Nonetheless, with a AGN wind velocity of $0.1c$, the average Eddington ratios are a factor of 2 - 3 lower than in the $v_w=0.03c$ case. The average Eddington ratios of the BHs in MW, Sbc and SMC are 0.04, 0.07 and 0.03 for $v_w=0.1c$ and 0.12, 0.15 and 0.06 for $v_w=0.03c$. The BHs accrete above 10\% of the Eddington limit for $\sim 8\%,10\%$ and $7\%$ of the simulated time with $v_w=0.1c$ in MW, Sbc and SMC, respectively. With the lower wind velocity this fraction increases to $\sim 22\%,18\%$ and $8\%$.

In Paper I we simulated the same three galaxies without AGN feedback. However, the Bondi accretion rate in those simulations had to be scaled down by a factor of $10^5$, owing to the extremely high nuclear densities in the absence of AGN feedback. With this factor of $10^5$ scaling, reasonable average Eddington ratios of 0.15 for MW, 0.13 for Sbc, and 0.11 for SMC were achieved in these no-feedback runs. In contrast, when using the AGN feedback model presented here, we find that the average central gas densities near the BH are about 3 orders of magnitude lower, and sound speeds are increased by about an order of magnitude. Consequently, AGN feedback reduces the average Bondi accretion rates by $\sim 5$ orders of magnitude, leading to BH growth rates similar to the scaled Bondi rates seen in Paper I. Therefore, in the presence of AGN feedback, we are able to obtain reasonable accretion rates with the Bondi prescription without any artificial scaling. (As noted in Section \ref{section:Methods}, in the no-AGN-feedback runs presented in this paper, the BHs are not allowed to accrete gas at all.)

\begin{figure*}
    \centering
    \includegraphics[width=\textwidth]{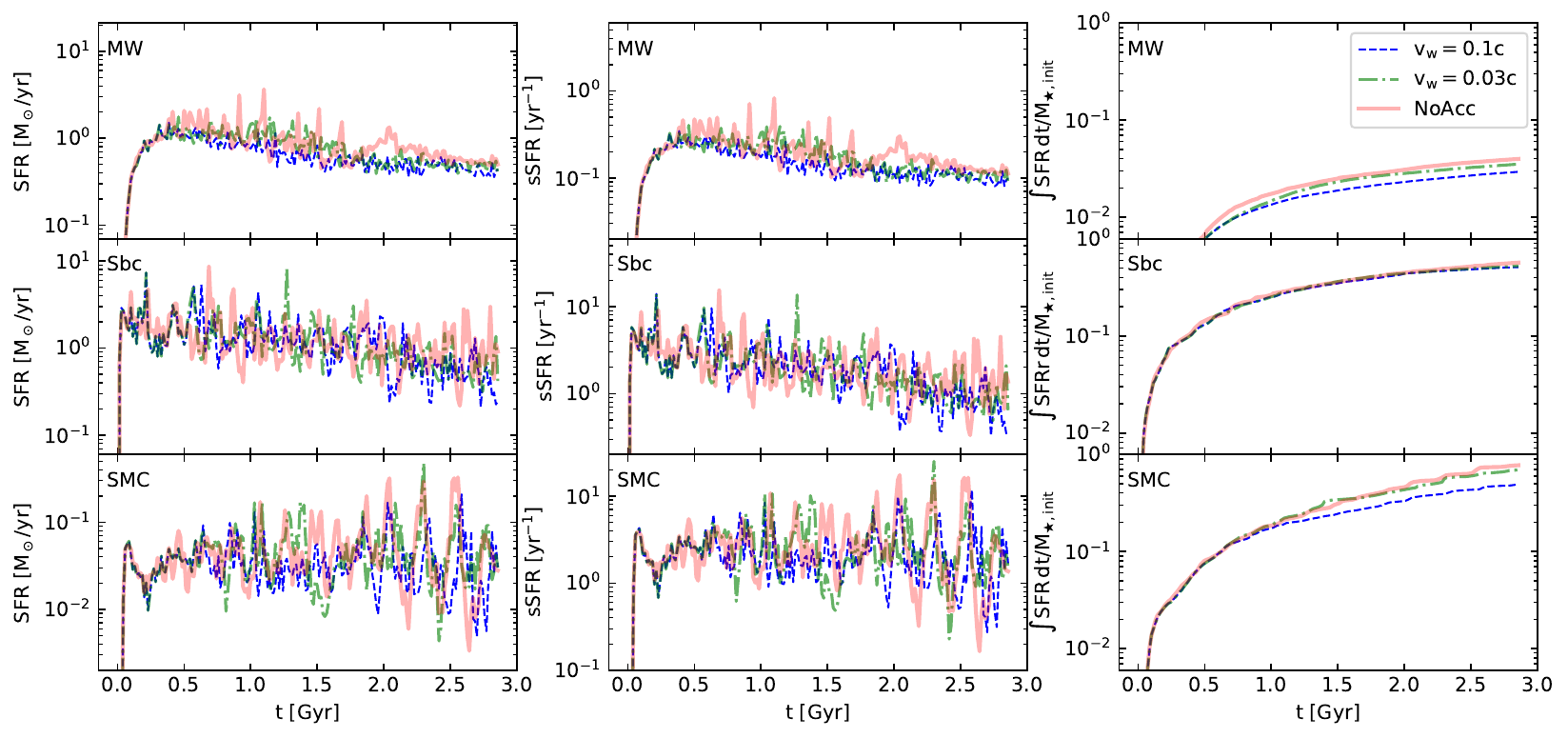}
    \caption{Total star formation rate (left), specific star formation rate (middle) and cumulative stellar mass formed normalized by initial stellar mass (right) in MW, Sbc and SMC runs with wind velocities set to 0.1$c$, $0.03c$ and without BH feedback. Note that the $y$-axis scales are different for each simulation in the left and middle panels. SMC and Sbc have more bursty star formation than MW. AGN feedback reduces the burstiness of star formation somewhat, which lowers the final stellar mass in the SMC and the MW galaxies, but has minimal impact on star formation in Sbc. 
    }
    \label{fig:SFR_Mstar}
\end{figure*}

\subsubsection{Mass outflow rates}
Next we look at the mass outflow rates in these simulations. We calculate the outflow rates using a shell method: we sum the radial momentum ($m_i v_{{\rm rad},i}$) of all the gas cells inside a thin shell centered around the BH and then divide it by the width of the shell ($\Delta r$), giving
\begin{equation}
    \dot{M}_{\rm out} = \frac{\sum_i m_i v_{{\rm rad},i}}{\Delta r} .
\end{equation}
It is also possible to calculate the outflow rate by integrating the radial flux across the surface of the sphere as $ \dot{M}_{\rm out} = \int \rho v_{\rm rad} dS$. We find that both methods give similar results in our simulations, and the shell method is computationally cheaper. We set the radius of the shell $(r_{\rm shell})$ to be 1.5 times the initial stellar half-mass radius ($r_{\rm half,\star}$) of the galaxies, which is $r_{\rm shell} = 10.3\,$kpc for MW, $3.73\,$kpc for Sbc and $2.61\,$kpc for SMC. We set the width of the shell to be $\Delta r = 0.71\,$kpc (= 0.5 kpc/h) for all the galaxies.  

In Figure \ref{fig:outflowrates} we compare the mass outflow rates of each galaxy, for wind velocities of 0.1c and 0.03c, as well as runs without AGN feedback. We show both instantaneous and cumulative outflow rates, each normalized by the initial gas mass within $1.5\,r_{\rm half,\star}$, which is $2.8\times10^9$\msun\ in MW, $2.1\times10^9$\msun\ in Sbc and $2.4\times10^8$\msun\ in SMC. In all three galaxies, the instantaneous mass flux oscillates between inflow and outflow over $\sim 100\,$ Myr timescales, even in the absence of AGN feedback. However, the magnitude of these fluctuations is significantly smaller in the MW than in the other two galaxies. The normalized peak outflow rates reach values of $\sim1$ Gyr$^{-1}$ in MW, 10-20 Gyr$^{-1}$ in Sbc, and $\sim40$ Gyr$^{-1}$ in SMC. 
(The corresponding non-normalized peak outflow rates are $\rm{\sim 3\;M_\odot yr^{-1}}$ in MW, $\rm{\sim 20-40\;M_\odot yr^{-1}}$ in Sbc and $\rm{\sim 10\;M_\odot yr^{-1}}$ in SMC.)  

The cumulative mass fluxes (integrated over the entire simulation length of 2.86 Gyr) on the right side of Figure \ref{fig:outflowrates} 
more clearly highlight qualitative differences between the three types of galaxies. The MW galaxy experiences a net inflow of gas through $r_{\rm shell}$, {\em regardless} of whether AGN feedback is included. AGN feedback reduces the average inflow rate (instantaneous flux averaged over the entire run)
from $0.1$ Gyr$^{-1}$ to $0.07$ Gyr$^{-1}$ or $0.02$ Gyr$^{-1}$ (for $v_w = 0.03$c \& $0.1$c, respectively). 
MW shows a clear trend of decreasing inflow with increasing feedback strength. The Sbc galaxy, in contrast, has a net {\em outflow} through $r_{\rm shell}$ in all cases, including the no-AGN-feedback run. In the two feedback runs the net flux becomes negative for very brief time intervals. There is no clear long-term impact of AGN feedback in the Sbc galaxy; the cumulative outflow is very similar in the runs with and without AGN feedback. This is because of the extreme bursty star formation in Sbc, as we will see below. 

The SMC galaxy presents still another qualitatively different scenario in which AGN feedback has a significant impact on the cumulative mass flux. In all three runs the cumulative flux has large oscillations. Without AGN feedback, the cumulative flux is mostly negative until $\sim2$ Gyr after which it has large oscillations around zero. Feedback with $v_{\rm w} = 0.03$c has minimal impact on the cumulative flux, which is negative throughout the run except a few brief intervals. (Note that these small differences are within the random fluctuations in the flux expected in bursty galaxies like Sbc and SMC.) In the feedback run with $v_{\rm w} = 0.1$c we do see some significant differences. The cumulative flux is positive throughout most of the simulation except during brief periods of time where it decreases slightly below zero. By the end of the run it reaches values as large as $\sim0.9$. The feedback also prevents the large inflows seen at later times in the no-feedback and $v_{\rm w} = 0.03$c runs.

\begin{figure}
    \centering
    \includegraphics[width=\columnwidth]{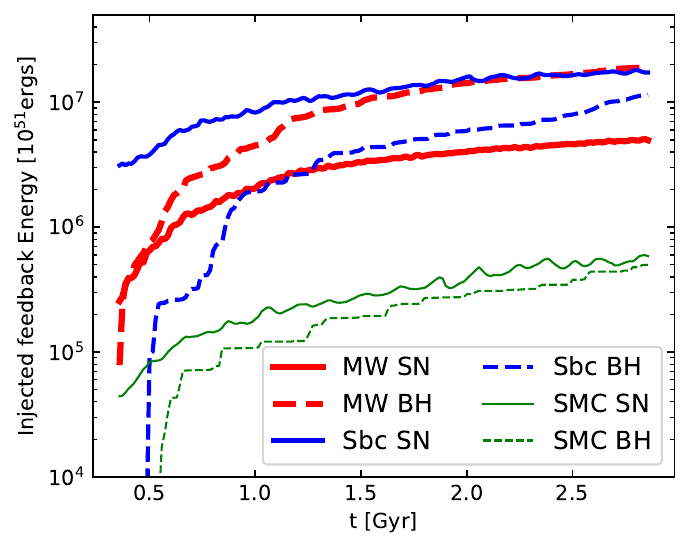}
    \caption{Cumulative SNII energy injected into the central 2 kpc region (solid lines) and cumulative BH feedback energy (dashed lines) in the MW, Sbc and SMC runs with $v_w=0.1c$. AGN feedback is subdominant to stellar feedback in Sbc, resulting in minimal impact of the AGN on this host galaxy. In the MW simulation, the AGN dominates the feedback energy budget, and in SMC AGN and stellar feedback energies are comparable.
    }
    \label{fig:SN_BH_energy}
\end{figure}

\begin{figure*}
    \centering
    \includegraphics[width=\textwidth]{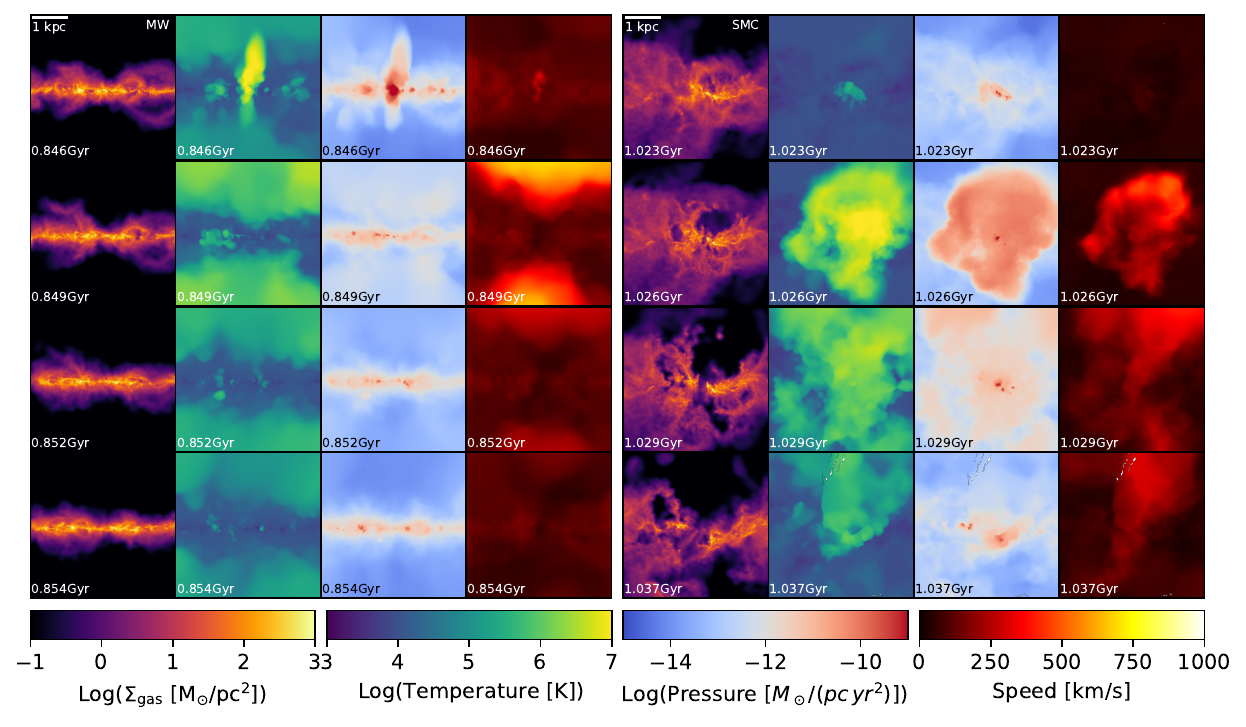}
    \caption{Edge-on projections of gas density (column 1), temperature (column 2), pressure (column 3) and speed (column 4) during a typical AGN outflow event in the MW (left 4$\times$4 block) and SMC (right 4$\times$4 block) galaxies with $v_w=0.1c$. Snapshot time increases from top to bottom (different for the two galaxies). A slice thickness of 1.4kpc is used for averaging the quantities. MW has a very thin disk which allows feedback to escape it with minimal work done resulting in low coupling efficiency. The thicker disk of SMC allows the feedback to do more work before leaving the disk resulting in more efficient coupling.}
    \label{fig:mw_smc_agn_2dplot}
\end{figure*}

\subsubsection{Star formation rates}
In Figure \ref{fig:SFR_Mstar} we compare the total star formation rates (SFRs), specific star formation rates (sSFRs), and the cumulative stellar mass (normalized by the initial stellar mass) formed in the three galaxies for $v_w=0.1c, \; 0.03c$ and in the runs without AGN feedback. The (s)SFRs highlight key differences between these three galaxies. The Sbc galaxy has a $\sim 10$ times higher sSFR than the MW, despite the fact that they have similar average SFRs (0.89 \msun\ yr$^{-1}$, 1.53 \msun\ yr$^{-1}$, respectively, in the absence of AGN feedback). Star formation is also much burstier in the Sbc galaxy than in the MW, as indicated by the larger stochastic variations. The SMC has similarly bursty star formation and a high sSFR, despite the much lower average SFR in this low-mass galaxy (0.059 \msun\ yr$^{-1}$). The high sSFRs in the Sbc and SMC galaxies reflect their higher initial gas fractions. 

The introduction of AGN feedback results in a slight decline in SFRs in the MW and SMC feedback runs after $\sim0.5$ Gyr. There is also a noticeable decline in the burstiness of star formation. However, in Sbc there is no significant change in SFR when AGN feedback is introduced. These effects of AGN feedback are clearer in the cumulative stellar mass plots on the right. In the SMC simulation, AGN feedback with $v_w=0.03c$ and $0.1c$ results in a decrease of 11\% and 37\% in stellar mass formed, respectively. In the MW simulation, star formation is decreased by 12\% for $v_w=0.03c$ and 26\% for $v_w=0.1c$. The impact in Sbc is even smaller. The cumulative stellar mass is roughly the same in all three runs until $t\sim 2$ Gyr. After this, the increased BH mass results in stronger feedback which decreases the cumulative stellar mass by 8\% and 10\% in the $v_w=0.03c$ and $v_w=0.1c$ runs, respectively. These trends align with the results of previous sections, and we conclude that the overall impact of AGN feedback on star formation is small to moderate in all three galaxies, but is especially weak in the Sbc galaxy.

\begin{figure*}
    \centering
    \includegraphics[width=\textwidth]{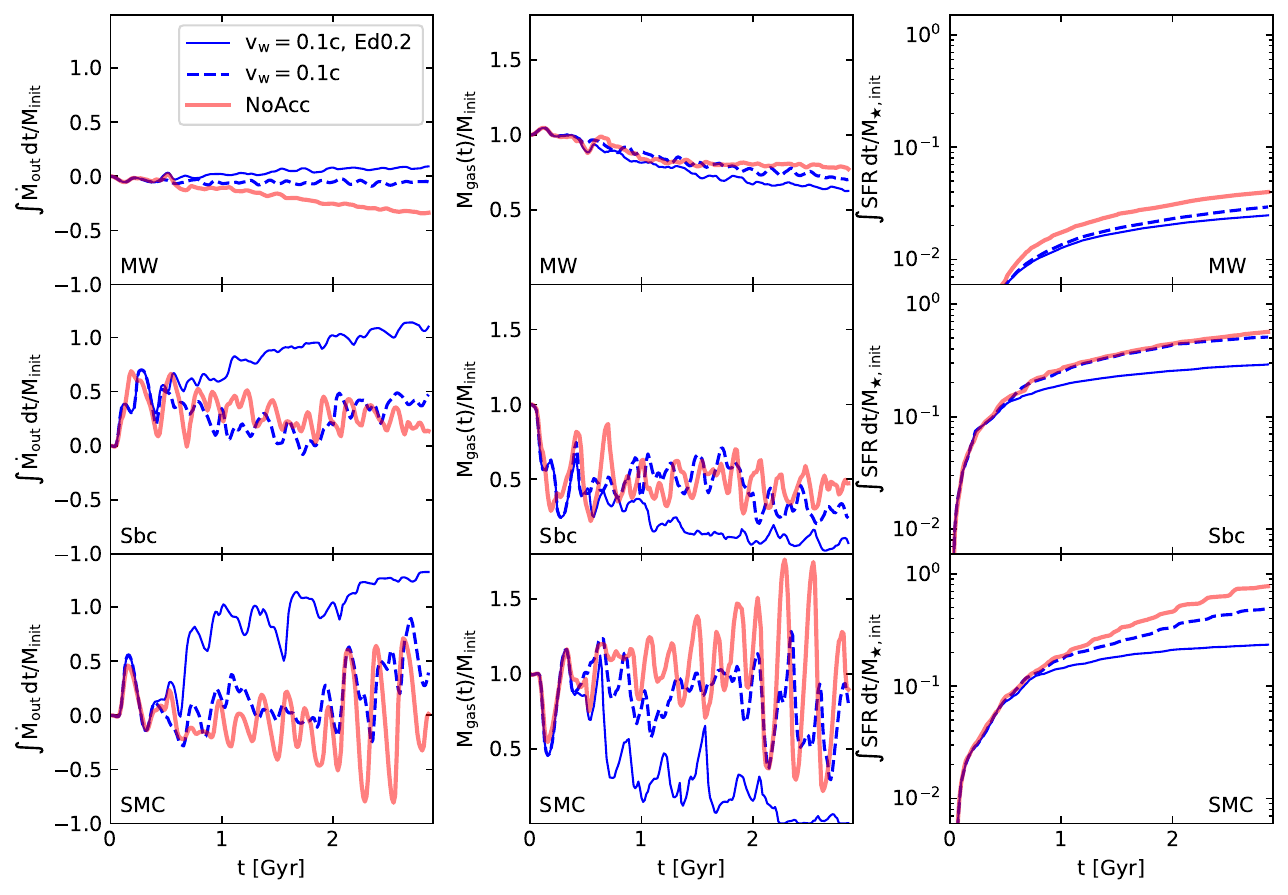}
    \caption{Host galaxy properties in simulations with the BH accretion rate artificially held to a constant value and a wind velocity of $0.1c$, compared to the fiducial $v_{\rm w} = 0.1c$ runs and the no-AGN-feedback runs. {\it Left:} Cumulative mass outflow rate normalized by the initial gas mass within 1.5$\times$ the stellar half mass radius for MW, Sbc and SMC runs. {\it Middle:} Gas mass within 1.5$\times$ the stellar gas mass radius normalized by the initial value for the same runs. {\it Right:} Cumulative stellar mass formed as a fraction of total initial stellar mass for the same runs. The higher accretion rates in the fixed-$\dot M$ runs lead to slightly higher outflows and lower star formation in MW. In contrast, SMC and Sbc shows much larger differences with most of the gas expelled by the end of the runs.}
    \label{fig:outflowratesEd10}
\end{figure*}

\subsubsection{Energetics and coupling strengths}
To better understand the physics behind these trends, in Figure \ref{fig:SN_BH_energy} we compare the energy injected into the central region of the galaxies by stellar feedback and by the AGN with $v_w=0.1$c. The dashed lines represent the cumulative energy injected by BH feedback, and the solid lines represent the cumulative SNII energy injected into the central 2 kpc region of the galaxies. Only in the MW does the AGN feedback energy dominate; by the end of the simulation, the AGN has injected several times more energy than stellar feedback. In contrast, stellar feedback energy consistently dominates over AGN feedback in both Sbc and SMC. This is especially true in the Sbc galaxy, which has strong, bursty star formation that is negligibly modulated by AGN winds.
In SMC, the AGN feedback energy is still subdominant but is only slightly lower than the stellar feedback energy. This results in a more noticeable impact of AGN feedback in the SMC than in the Sbc galaxy.

Despite the fact that the AGN is energetically dominant in the MW and is subdominant in the SMC, we saw in the previous subsections that star formation is more strongly suppressed by AGN feedback in the SMC, at least for the $v_{\rm w} = 0.1c$ case. This is primarily due to the different coupling efficiencies of the feedback in the two systems. To illustrate this, Figure \ref{fig:mw_smc_agn_2dplot} shows the edge-on view of the gas temperature, pressure, density and velocity in the central 4 kpc region of the MW and SMC galaxies during a typical AGN outflow event in each galaxy. By comparing the initial gas density plots of MW and SMC, we see that MW is a thin and concentrated disk compared to SMC, which is thicker and more diffuse. This allows the feedback winds in MW to escape more easily in the vertical direction without doing much $PdV$ work on the gas disk. During the initial stage, the hot pressurized gas expands mostly in the vertical direction, along the path of least resistance. After leaving the disk plane it expands in the horizontal directions. There is only a little impact on the dense gas in the disk plane. 

However, in SMC, the hot pressurized bubble of gas created by the AGN feedback expands in all three directions inside the disk plane to a larger volume before escaping in the vertical direction. It also has much more impact on the dense gas in the disk. In the third snapshot of the gas density distribution, we can see that the feedback ejects some of the dense gas above and below the BH, completely disrupting the morphology of gas in the central region around the BH. Thus, the diffuse vertical structure of the gaseous disk of SMC allows a more efficient coupling of the AGN feedback to the galaxy, whereas the thinner disk of MW results in a weaker coupling. We also note that the potential well of SMC is much shallower than that of the more massive MW and Sbc galaxies. This factor also contributes to the larger outflows detected in SMC. 

\subsection{Simulations with fixed accretion rate}
\label{ssec:fixed_mdot}
The outflow rates and quenching effects of AGN feedback on host galaxies are influenced by the surrounding ISM conditions, which in turn affect both the BH accretion rates and the AGN feedback coupling strength. Decoupling these two factors is very challenging due to their highly nonlinear interdependence. To isolate and compare the effects of feedback coupling to the ISM, we simulated the same three galaxies with BH accretion rates set to 20\% of the initial Eddington accretion rate throughout the simulation and the wind speed set to 0.1c. We choose a slightly higher accretion rate than the average Eddington ratio of the fiducial runs in order to increase the visibility of feedback effects, while still avoiding artificial effects such as persistent cavities around the BH due to extreme feedback injection.

\subsubsection{Outflow and star formation rates}
 In the leftmost column of Figure \ref{fig:outflowratesEd10}, we show the cumulative mass outflow rates normalized with respect to the initial gas mass within $1.5\times r_{\rm half,\star}$ of each galaxy for the fixed-$\dot M$ runs, compared to the corresponding fiducial runs with and without AGN feedback. We also show the normalized gas masses within the same radii in the middle column of Figure \ref{fig:outflowratesEd10}. With the increased accretion rates, the feedback energy budget is dominated by the AGNs in all three galaxies. In MW, the increased accretion rate only causes a marginal increase in the impact of feedback. The average normalized outflow increases to 0.03 Gyr$^{-1}$ compared to -0.01 Gyr$^{-1}$ in the fiducial run and -0.1 Gyr$^{-1}$ in the no-feedback run.  (Recall that these rates are normalized by the initial gas mass inside $1.5\times r_{\rm half,\star}$). 
 
 AGN feedback has an even more dramatic effect in the fixed-$\dot M$ Sbc and SMC runs, relative to their fiducial counterparts. In Sbc, the fiducial simulations had modest cumulative mass outflows with and without AGN feedback. In the fixed-$\dot M$ Sbc simulation, however, we see large outflows with the cumulative mass outflow rate more than twice that of the no-AGN-feedback run. By the end of the simulation almost all of the gas in the central region has been evacuated by the feedback. (Note that the normalized, cumulative mass outflow in the fixed-$\dot M$ Sbc and SMC runs exceeds unity at the end of those simulations, which results from sampling errors due to the lack of enough gas cells.) In the case of SMC, the fiducial feedback run suppressed the inflows and produced a net outflow. With the higher (constant) accretion rate, we see much stronger net outflows, which are also larger than that of the Sbc fixed-$\dot M$ run. Similar to the case of Sbc, essentially all the gas in the central region of the SMC is evacuated by the feedback. This underscores our finding in \S~\ref{ssec:fiducial} that increasing the AGN feedback strength will result in larger outflows in thicker disks due to the high coupling efficiency, especially in low-mass galaxies, while in thinner disks, the increase in outflows will be more modest.

In the rightmost column of Figure \ref{fig:outflowratesEd10} we show the cumulative stellar mass formed normalized by the initial stellar mass as a function of time in the same galaxies. In fixed-$\dot M$ case, the suppression in star formation due to AGN feedback is stronger in relation to their fiducial counterparts, with the largest effect seen for the SMC. Relative to the no-feedback runs, star formation is lowered by 38\% in MW, 49\% in Sbc and 70\% in SMC for these fixed-$\dot M$ runs. In  comparison, the fiducial AGN feedback suppressed star formation by 26\% in MW, 10\% in Sbc, and 37\% in SMC, respectively, relative to the no-feedback case. Thus the star formation rates are much more sensitive to the accretion rates in Sbc and SMC compared to MW, which is similar to the trend seen in the outflow rates. These two trends again imply that the feedback coupling is much stronger in Sbc and SMC compared to MW. 

Overall, these findings reinforce a major conclusion from the fiducial simulations: namely, that the impact of AGN feedback depends strongly on the morphology and nuclear environment of the host galaxy. In both the fiducial and fixed-$\dot M$ runs, the SMC galaxy has the strongest coupling of AGN feedback to the surrounding ISM. This trend seen in the fiducial simulations persists when the accretion rate is fixed to a constant value demonstrates that the differences in host environment, rather than self-regulation of BH fueling, are driving the differences in AGN feedback coupling.

\begin{figure*}
    \includegraphics[width=0.95\textwidth]{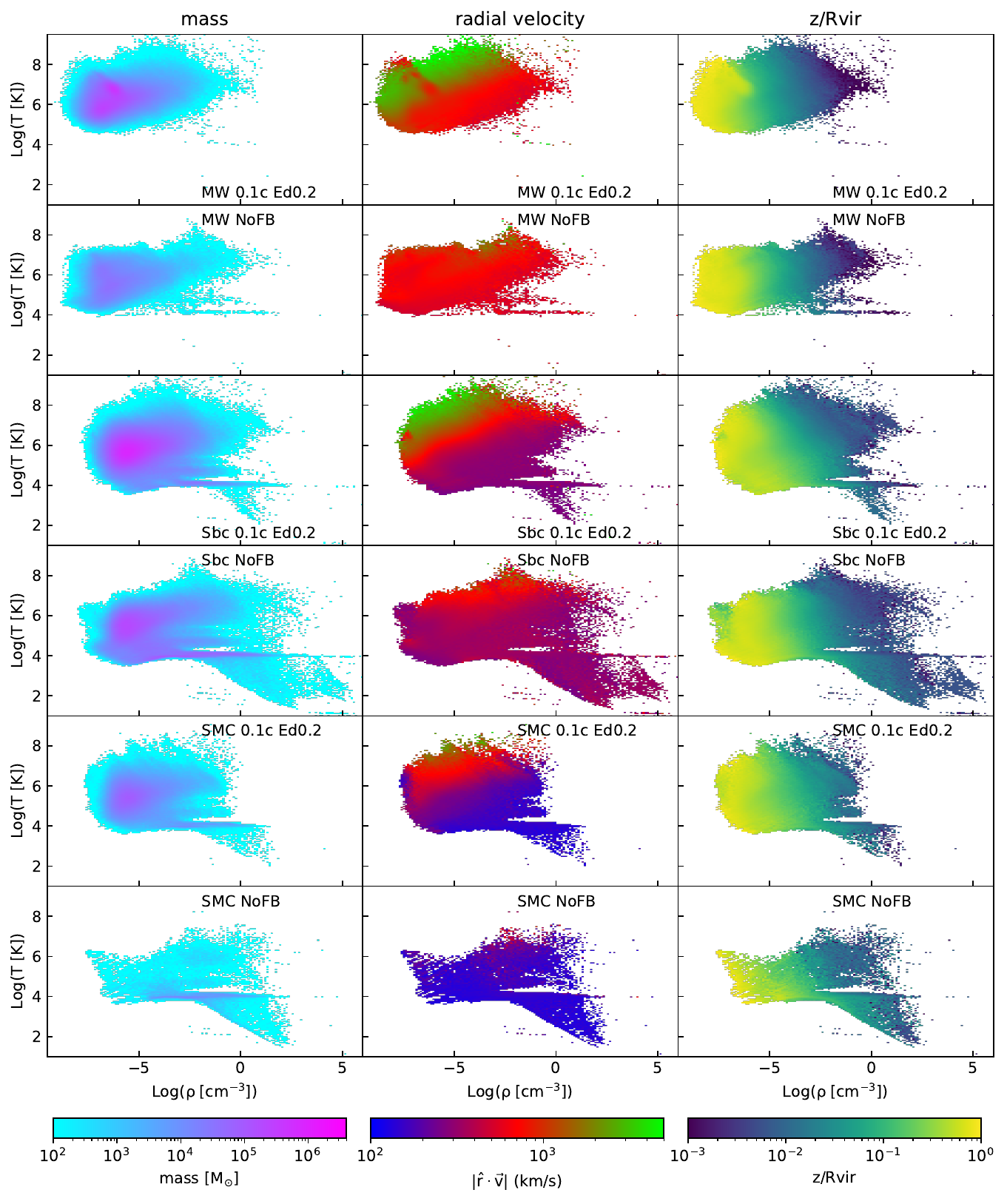}
    \caption{Temperature - Density phase space diagram  of the outflowing gas averaged over the entire simulation and color coded by mass (column 1), mass-weighted outward radial velocity (column 2) and mass-weighted distances in the vertical direction scaled by the disk height (column 3) in MW, Sbc and SMC runs without feedback and with $v_w =0.1c$ at t=2 Gyr. Only gas cells with outward radial velocity $>$ 3 times the virial velocity and located within the virial radius are shown. All three feedback runs have more outflowing gas than the corresponding no-feedback runs. Runs with AGN feedback have bimodal distributions separated by 1-2 orders of magnitude in temperature. MW produces the highest outflow velocities. Feedback heats up the gas in all three galaxies with largest shift in SMC and smallest shift in MW. SMC and Sbc also have more gas within the disk compared to MW. 
    }
    \label{fig:phasespace}
\end{figure*}

\subsubsection{Phase space structure}
\label{ssec:phase space}
The strong impact of AGN feedback in the fixed-$\dot M$ simulations allows us to identify key differences in the gas phase-space structure in galaxies with and without AGN feedback. In Figure \ref{fig:phasespace}, we compare the phase space properties of the high-velocity outflowing gas in the three galaxies with fixed $\dot M$ and without AGN feedback. The phase space plots are averaged over the entire simulation ($t > 0.36\,$Gyr) and the bins in $T-\rho$ space are color-coded in each panel to indicate the average mass, the mass-weighted average radial velocity, and the average vertical distance in units of disk scale height. To separate outflowing gas from the bulk motion in the galaxy, we consider only high-velocity outflowing gas, defined as gas cells with radial outward velocity greater than 3 times $v_{200}$ of the system. The corresponding velocity cutoffs are 452 km/s for MW, 245 km/s for Sbc and 123 km/s for SMC. We also plot only gas cells that are within the virial radius of the galaxies to avoid boundary effects at the edges of the simulation box.

From the plots in the first column we can see that most of the rapidly outflowing gas falls within the temperature range of $10^4$K to $10^8$K and has densities between $10^{-8}\mathrm{cm}^{-3}$ and $10^{-3}\mathrm{cm}^{-3}$, with a higher-density tail. The only exception is the SMC run without feedback, which has very little outflowing gas at all. In each set of simulations, those with and without feedback, Sbc has the most outflowing material and SMC has the least. 

As expected, the runs with AGN feedback have more outflowing gas  at higher temperatures, as well as more massive outflows overall. The total outflowing gas mass is greater by factors of 4.8, 2.6 and 14 in the MW, Sbc and SMC  runs with feedback relative to their counterparts without feedback. (Note that these numbers will vary with the adopted velocity cutoffs.) The mass-weighted median temperature of the outflowing gas also increased by factors of 3.9, 2.7 and 28 in MW, Sbc and SMC relative to the no-feedback runs. These numbers are consistent with the trends seen in the previous section, namely that the strongest AGN feedback coupling occurs in the SMC galaxy.

The most noticeable difference between the distributions in the feedback vs no-feedback runs is observed in the low density ($<10^{-3}\mathrm{cm}^{-3}$) region of the phase space, which is significantly more occupied in the feedback runs. These gas cells constitute $>95\%$ of the outflowing gas in all three galaxies. From the velocity and $z$ color coding we can see that most of these gas cells are located far ($\sim\mathrm{R_{vir}}$) from the galactic plane. These cells are also flowing out with velocities significantly higher than that of the cells in the same phase space region of the corresponding no-feedback runs. The feedback runs also feature some outflowing gas with temperatures exceeding $10^8$ K, which also possess the highest velocities within the distributions. This portion of the phase space is nearly vacant in the no-feedback runs. These signatures indicate that these gas cells were shock heated by AGN feedback.

From the plots in the second column we can see that the highest radial velocities are obtained for outflowing gas in the MW feedback run. Both MW and Sbc feedback runs have significant amounts of gas with velocities of the order of several thousands of km/s, but MW has roughly $\sim8$ times more gas with velocities $> 4000$ km/s. Note that both galaxies have very similar initial gas content. Because of the thin disk of the MW, feedback does work on a much smaller amount of the ISM gas compared to that of Sbc, allowing the outflowing gas to retain high velocity. In SMC, there is only a negligible amount of gas with these velocities. 
 
By comparing the vertical distribution of the gas between the three feedback runs in the third column, we find that around 0.06\%, 0.8\% and 1.4\% of the outflowing gas in MW, Sbc and SMC, respectively, is located within $\sim4$ times the disk scale height $(|z|<2h)$. This again suggests that most of the feedback energy in MW escapes into the vertical direction, whereas in Sbc and in SMC, relatively more work is being done within the disk plane (with the largest impact in SMC).

By comparing the three feedback runs we see that MW has no outflowing material in the atomic or molecular phase (expected with T$<10^4$K and $\rho>10^{-1}\mathrm{cm^{-3}}$) whereas Sbc and SMC feature some material in this phase. This is somewhat expected since a higher disk thickness makes it easier for the feedback to entrain cold and dense gas.

\begin{figure*}
    \includegraphics[width=\textwidth]{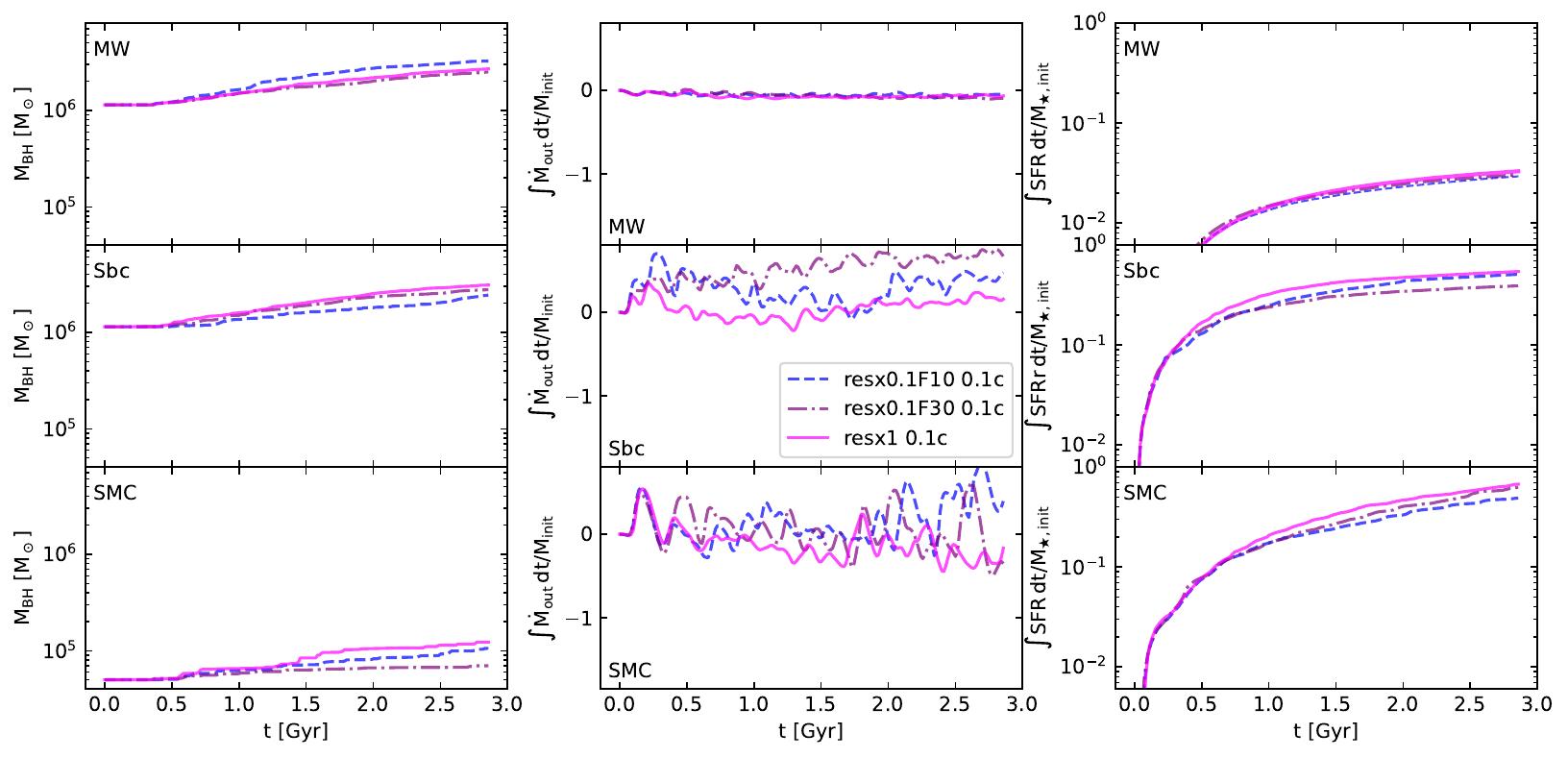}
    \caption{Results of resolution convergence tests are shown for the fiducial MW, Sbc, and SMC galaxies with the $v_w = 0.1c$ AGN feedback model. The \textit{left} column shows BH mass growth in the same manner as Figure \ref{fig:bhmass_eddratio_vs_vel}, the \textit{middle} column shows the cumulative mass outflow rate normalized to the initial gas mass within $1.5\times$ the stellar half-mass radius, in the same manner as Figure \ref{fig:outflowrates}, and the \textit{right} column shows the cumulative stellar mass formed, normalized to the total initial stellar mass, in the same manner as Figure \ref{fig:SFR_Mstar}. In each column, panels show results for the MW (\textit{top}), Sbc (\textit{middle}) and SMC (\textit{bottom}) galaxies, each simulated with the fiducial resolution (resx0.1F10, dashed lines), with a $3\times$ higher resolution near the BH (resx0.1F30, dot-dashed lines), and with a resolution throughout the whole simulation volume that matches the fiducial resolution near the BH (resx1, solid lines). The final BH mass varies by $\sim 50\%$ for different resolution simulations of each galaxy. Outflows and star formation rates are well converged in the MW runs, but the SMC and Sbc galaxies show somewhat poorer convergence due to the stochasticity induced by extreme bursty star fomation.}
    \label{fig:resolution convergence}
\end{figure*}

\subsection{Resolution Convergence}
\label{ssec:resolution convergence}
In Figure \ref{fig:resolution convergence} we analyze the resolution convergence of three quantities: BH mass growth, cumulative (normalized) mass outflow rates and cumulative (normalized) star formation rates in the three galaxies. We compare the fiducial resolution runs (resx0.1F10) with ${v_{\rm w} = 0.1c}$ to runs with a uniform resolution (resx1) that is the same as the central resolution of the fiducial run, as well as runs with same background resolution but with a 3 times larger refinement factor (resx0.1F30). Because these additional simulations alternately vary the background and the central resolution, relative to the fiducial setup, these comparisons do not constitute a study of the rate of convergence. Nonetheless, they serve to demonstrate that our conclusions are robust to variations in simulation resolution.

We first note that, while variations in these quantities with resolution do exist, the variations do not exhibit discernible systematic trends. Some variation is expected between different realizations of similar initial conditions, especially given the highly nonlinear and stochastic nature of the explicitly resolved, multiphase ISM as well as the $M_{\rm BH}^2$ scaling of the Bondi accretion rate. Accordingly, we see that MW shows the best resolution convergence among the three galaxies. This can be attributed to the relatively weak stellar feedback in MW, leading to reduced stochasticity and randomness within the ISM. The BH mass growth, which again is notoriously sensitive to small changes in initial conditions within a Bondi accretion scenario, exhibits $\sim50\%$ variation by the end of the MW simulation. In contrast, the cumulative mass outflow rate and stellar mass formed have almost no variations between the different resolutions. 

The BH mass growth curves in Sbc and SMC have fluctuations with similar magnitudes to those in MW. However, because Sbc and SMC have much more bursty star formation than MW as shown in Figure \ref{fig:SFR_Mstar}, these galaxies are more susceptible to stochastic fluctuations in the gas dynamics. As a result, the integrated outflow and star formation rates show significantly more fluctuation between the different resolution simulations, with roughly $\sim100\%$ and $30\%$ spread respectively in both galaxies. The normalized, cumulative flux in SMC switches from a slight net outflow of 0.14 Gyr$^{-1}$ in the fiducial run to slight net inflows of 0.12 Gyr$^{-1}$ in the resx0.1F30 run and 0.06 Gyr$^{-1}$ in the resx1 run. In the case of Sbc, all three resolutions show net outflows. However, the net flux changes from 0.17 Gyr$^{-1}$ in the fiducial run to 0.23 Gyr$^{-1}$ in the resx0.1F30 run and 0.05 Gyr$^{-1}$ in the resx1 run. We again emphasize that no systematic trend is seen in these variations with resolution. Thus, although our quantitative results are susceptible to these fluctuations with different resolutions, the qualitative results from the comparisons between the feedback vs. no-feedback runs and the different ICs remain minimally affected.

\section{Discussions and Conclusions}
\label{section:discussions&conclusions}
In this paper we modeled AGN feedback in the form of fast nuclear winds in AREPO hydrodynamics simulations of idealized isolated galaxies. Our simulations utilized the explicit ISM and stellar feedback model SMUGGLE, which generates a multiphase ISM and implements localized stochastic star formation and stellar feedback. We simulated three types of galaxies: a MW-type, gas-poor galaxy with a thin disk, a gas-rich, SMC type galaxy with a thick gas disk, and a LIRG-like galaxy (Sbc) that is also gas rich and has a thick gas disk. Using a super-Lagrangian refinement scheme, we were able to resolve gas dynamics at $\sim10-100$ pc scales in the vicinity of the BHs.

Our simulations show that the AGN wind feedback is very efficient at regulating the BH mass growth in all three galaxies. In the absence of AGN feedback, the cold and dense gas clumps in the central region of the galaxies result in peak Bondi accretion rates that are many orders of magnitude above the Eddington limit \citep{sivasankaran2022simulations}. Our wind feedback mechanism results in lower gas densities and higher temperatures around the BHs, producing reasonable accretion rates with the Bondi prescription --average Eddington ratios in our fiducial simulations range from 0.03 to 0.15, with lower Eddington ratios corresponding to higher AGN wind velocities. Over a period of 2.5 Gyr, the BHs grow by factors of 2-3 with a wind velocity of 0.1c. When the wind velocity is lowered to $0.03$c, the BH mass in MW and Sbc increases by factors of $\gtrsim 10$ over the same period. In SMC, which has a smaller initial BH mass, BH growth is similarly enhanced with weaker AGN feedback but to a lesser degree, owing to the $M_{\rm BH}^2$ scaling of the Bondi accretion rate. 

Despite the clear long-timescale impacts of AGN feedback on BH fueling, we find that the BH accretion rate still fluctuates by orders of magnitude on short ($<100$ Myr) timescales, as in \citep{sivasankaran2022simulations}. The characteristic variability timescales appear to be independent of AGN feedback strength and are instead dominated by rapid variations in the central gas density, owing to stellar-feedback-driven turbulence in the ISM.

Turning to examine the impact of AGN feedback on the host galaxy, we find significant qualitative differences between the three types of simulated galaxies. The SMC experiences the strongest AGN-driven outflows and suppression of star formation, with a 37\% lower mass in newly formed stars in the $v_{\rm w}=0.1c$ case compared to the no-AGN-feedback run. In the MW galaxy, this high-velocity AGN feedback led to a 26\% decline in new star formation. Unlike the SMC, however, the MW galaxy does not experience large fluctuations between massive inflows and outflows, and the cumulative mass flux through the nuclear region is a net inflow throughout the simulations, even in the presence of strong AGN feedback. Large variations in central mass flux are also seen in the Sbc run, but the impact of AGN feedback is more modest: only 10\% fewer new stars form with $v_{\rm w}=0.1c$ AGN feedback compared to the no-AGN-feedback case.

We find that these trends are controlled by two main factors: the relative strength of AGN vs. stellar feedback in the galactic nucleus, and the host galaxy morphology (specifically, the thickness of the gas disk and the depth of the central potential). In the MW, which has relatively quiescent star formation, the AGN energy input easily dominates over the stellar feedback energy in the galactic nucleus, leading to a significant suppression in star formation. However, the MW has a very thin disk, which allows the much of the AGN feedback to escape easily through the path of least resistance in the vertical direction with minimal work being done on the gas disk. Thus, the cumulative mass flux remains a net inflow, even in the presence of energetically dominant AGN feedback. This result is in agreement with the findings of \citealt{Torrey2020} using the Feedback in Realistic Environments (FIRE) framework. They also found that AGN feedback couples inefficiently to thin disc galaxies, with a significant fraction of the energy venting in the polar directions.

Contrary to MW, the thick and diffuse gas disks of Sbc and SMC allow for a much more efficient coupling of AGN feedback to the ISM. Feedback does more work inside the disk during the initial expansion phase before escaping from the disk plane. At the same time, the Sbc and SMC also have much burstier star formation than the MW, and the resulting stellar feedback also couples more efficiently to their thicker gas disks than in the MW. In the Sbc galaxy, AGN feedback energy is subdominant to stellar feedback in the nuclear region; thus, we see minimal impact of AGN feedback on star formation and gas outflow rates, despite its efficient coupling. In the SMC simulations, meanwhile, the total energy injected by the AGN versus stellar feedback in the central region is comparable. The SMC is also a low-mass galaxy with a shallower potential, from which outflowing gas can more easily escape. These factors combined with the efficient feedback coupling in the SMC galaxy yield the strongest AGN-driven outflows and star formation quenching among the three galaxy types considered.

To further isolate the effects of the host galaxy environment on the impact of AGN feedback, we simulated the three galaxies with fixed BH accretion rate and with the AGN wind velocity set to $0.1c$. This exercise effectively eliminates the non-linear dependence of AGN feedback coupling on the BH accretion rates from our analysis. By choosing a fixed $\dot M$ of 20\% of the initial Eddington rate, which is somewhat higher than the average Eddington ratio in our fiducial runs, we also amplify the impact of AGN feedback to make trends more readily apparent. In MW the effects are modest, changing the cumulative central mass flux from a slight inflow to a slight outflow and decreasing new star formation by 38\% relative to the no-AGN-feedback (versus 26\% with fiducial AGN feedback). In Sbc and SMC, however, the
higher, fixed accretion rate and efficient coupling leads to strong outflows and quenching effects. Here the AGN feedback in the Sbc finally becomes energetically dominant, and the impact on new stellar mass formed is five times larger than in the fiducial run. Little gas remains in the central region of the galaxy by the end of the run. Regardless, the SMC again experiences the strongest impact from AGN feedback in these fixed-$\dot M$ runs, with 70\% less star formation than in the no-AGN-feedback run, and almost no gas remaining in the central region of the galaxy at the end of the run. The changes in the outflow and star formation rates with the fixed-$\dot{M}$ are consistent with our findings from the fiducial simulation, further validating our conclusions. 

Our phase space analysis of the fast outflowing gas in the fixed-$\dot{M}$ runs also revealed similar trends. The fraction of outflowing material located within the gas disk was much higher in SMC and Sbc compared to MW, indicating more work done on the ISM by AGN feedback in these galaxies. Additionally, we find that the outflows generated by the AGNs are mainly made up of hot and low density gas in all three galaxies. In Sbc and SMC galaxies we found small amounts of gas with temperatures and densities corresponding to atomic and molecular phase. Outflows in MW did not have any gas in the atomic and molecular phases. The phase space distributions of the feedback runs also exhibited signatures of shock heated gas with very high temperatures $(>10^8\mathrm{K})$ and velocities ($>1000\mathrm{\,km\,s^{-1}}$) which were absent in the outflows generated by stellar feedback alone. 

In summary, we have studied the impact of AGN feedback in a set of three isolated galaxies with varying AGN wind velocities. While the AGN are highly variable within the stochastic, multiphase ISM, they have moderate average Eddington ratios. We avoid simulating the regime of massive elliptical galaxies with extreme quenching, where AGN feedback is expected to have strong global effects on the host \citep[e.g.,][]{springel2005black,dubois2013agn}. Instead, we deliberately focus on a small suite of isolated galaxies to probe in detail the interplay of BH fueling, AGN feedback, star formation, and stellar feedback, in the absence of extreme quasar phases or secular effects such as galaxy interactions. These simulations reveal efficient self-regulation of BH growth and a complex picture of BH/galaxy co-evolution. Our investigation reveals that AGN luminosity and wind velocity, gas disk morphology, host potential, and stellar feedback strength are the key factors determining the AGN-galaxy coupling. In certain regimes, specifically in galaxies with thick gas discs and luminous AGNs, the feedback can significantly reduce star formation and alter the gas content and morphology of the host. 

In forthcoming work, we extend our analysis to include merging galaxies. Galaxy mergers are thought to be important drivers of BH/galaxy co-evolution. They can trigger bursts of star formation and AGN activity by rapidly funneling gas into galactic nuclei \citep[e.g.,][]{sanders1990ultraluminous, hernquist1989, barneshernquist1991, barneshernquist1996, mihoshernquist1996}. The resulting feedback can quench star formation, while the merging galaxies simultaneously undergo morphological transformations \citep{springel2005modelling,hernquist1992structure,hernquist1993structure}. However, the extent of the connection between galaxy mergers and AGN fueling remains a subject of debate, underscoring the need for improved constraints from simulations \citep[e.g.,][]{satyapal2014galaxy,weston2016incidence,goulding2018galaxy,Grogin2005,Pierce2007,villforth2019host}. The ability to explicitly resolve localized, stochastic star formation and a multiphase ISM with models like SMUGGLE has opened a new discovery space for studies of BH/galaxy co-evolution. Our findings in this work highlight the rich and complex physics that such studies will continue to reveal.

\section*{Acknowledgements}
LB acknowledges support from NASA Astrophysics Theory Program awards 80NSSC20K0502 and 80NSSC22K0808. The simulations in this study were performed using the supercomputing cluster HiPerGator at University of Florida.

\section*{Data Availability}
The data underlying this article will be shared on reasonable request to the corresponding author.



\bibliographystyle{mnras}
\bibliography{references} 








\bsp	
\label{lastpage}
\end{document}